\documentclass[a4paper,12pt,dvips]{article}   
\usepackage{amsmath,amssymb,exscale}   
\usepackage{array}   
\usepackage{afterpage,float,flafter}   
\usepackage{epsfig,rotating,pifont,fancybox}   
\newcommand{\bb}{\bigskip}   
\makeatletter   
\@addtoreset{equation}{section}   
\makeatother

\catcode`\@=11   
\def\marginnote#1{}   
\newcommand{\bea}{\begin{eqnarray}}   
\newcommand{\eea}{\end{eqnarray}}   

\newcommand{\NPB}[3]{\emph{ Nucl.~Phys.} \textbf{B#1} (19#2) #3}   
\newcommand{\PLB}[3]{\emph{ Phys.~Lett.} \textbf{B#1} (19#2) #3}   
\newcommand{\PRD}[3]{\emph{ Phys.~Rev.} \textbf{D#1} (19#2) #3}

\newcommand{\MPL}[3]{\emph{ Mod.~Phys.~Lett.} \textbf{A#1} (19#2) #3}   
\newcommand{\PR}[3]{\emph{ Phys.~Rep.} \textbf{#1} (19#2) #3}

\newcommand{\Tr}{\mathop{\rm Tr}}

\relax   
\def\simlt{\stackrel{<}{{}_\sim}}

\title{   
\vspace{-0.7cm}   
\begin{flushright}   
\normalsize{   
CERN-TH/98-393\\   
IEM-FT-186/98 \\   
hep--ph/9812489\\ 
}   
\end{flushright}    
\vspace{1cm}   
\Large\textbf{Supersymmetry and Electroweak breaking from extra    
dimensions at the TeV-scale~\footnote{Work supported in part by the CICYT of  
Spain (contracts AEN98-0816 and AEN95-0882).}}}   
\vspace{1cm}   
\author{A.~Delgado~$^a$, A.~Pomarol~$^b$\footnote{On leave from:    
IFAE, Universitat Aut{\`o}noma de Barcelona, E-08193 Bellaterra, Barcelona.}\,    
and M.~Quir{\'o}s~$^a$\\   
\vspace{.5cm}\\   
$^a$~Instituto de Estructura de la Materia (CSIC),    
Serrano 123,\\   
 28006-Madrid, Spain\\   
\vspace{.2cm}\\   
$^b$~TH-Division, CERN,   
CH-1211 Geneva 23, Switzerland}   
\date{}   
\begin{document}   
\maketitle   
\begin{abstract}   
We analyze some features of the r{\^o}le that extra dimensions, 
of radius $R$ in the TeV$^{-1}$ range, can play in the  
soft breaking of supersymmetry and the spontaneous breaking of electroweak   
symmetry. We use a minimal model where the gauge and Higgs sector of the 
MSSM are living in the bulk of five dimensions and the chiral multiplets in a   
four-dimensional boundary. Supersymmetry is broken in the bulk by the 
Scherk-Schwarz mechanism and transmitted to the boundary by radiative  
corrections. The particle spectrum is completely predicted as a  
function of a unique $R$-charge. 
The massless sector corresponds to the pure Standard Model  
and electroweak symmetry is radiatively broken with a light 
Higgs weighing $\simlt$ 110 GeV. The $\mu$-problem is solved and Higgsinos,  
gauginos and heavy Higgses acquire masses $\sim 1/R$. Chiral 
sfermions acquire radiative squared-masses $\sim \alpha_i/R^2$.  
The effective potential is explicitly computed in the  
bulk of extra dimensions and some cosmological consequences can be  
immediately drawn from it. Gauge coupling running and 
unification is studied in the presence of Scherk-Schwarz 
supersymmetry breaking. The unification is similar to that in the 
supersymmetric theory. 
\end{abstract}   

\vskip1.cm   
   
\begin{flushleft}   
IEM-FT-186/98\\   
CERN-TH/98--393\\   
December 1998 \\   
\end{flushleft}   
\newpage   
\section{Introduction}   
The Standard Model (SM) of strong and electroweak interactions 
is being tested by present high-energy colliders for energies 
$\simlt$ 200 GeV and proved to describe the corresponding  
interactions with great accuracy. By the same token those 
experiments are putting limits on the scale of new physics, 
suggesting thus that extra matter, if it exists, is only relevant in the 
TeV scale range. In particular these limits do apply to the best 
motivated of the phenomenological low-energy extensions of the 
SM, its minimal supersymmetric extension (MSSM), putting bounds 
around the TeV on the mass of the supersymmetric partners. On the other 
hand, the origin of the electroweak symmetry breaking (EWSB) 
mechanism remains as the missing building block of the SM, or any low 
energy extension thereof, given the fact that the Higgs field has been 
shown so elusive in all experiments to the present day. In the 
context of the MSSM the problems of the origin and stability of 
the EWSB scale are alleviated because the latter is related to 
the scale of supersymmetry breaking, and it is further protected 
against radiative corrections by supersymmetry: the sensitivity 
of the Higgs squared-mass to the high scale is not quadratic (as in the 
case of the SM) but only logarithmic. However the origin of 
supersymmetry breaking remains as the big unsolved problem in 
these theories. 
  
The presence of supersymmetry in the low-energy extension of the 
SM is further supported by the fact that consistent  
fundamental theories aiming to 
unify gauge and gravitational interactions at high scales 
(string theories) are supersymmetric. A common feature of these  
theories is the presence of (compactified) extra dimensions. If 
all dimensions are small, of the order of the Planck or 
GUT length, their detectability is outside the scope of 
present or future accelerators. However if some radii are 
larger, they might have a number of theoretical and 
phenomenological implications~\cite{a}-\cite{aq1}.  
 
From the fundamental point of view the presence of large extra 
dimensions has been proved to be essential to describe the strong coupling 
regime of certain string theories~\cite{strong}, while TeV scale superstrings 
have been constructed~\cite{TeV,TeV2,zurab}.  
From the more phenomenological point of 
view the presence of large dimensions can play a prominent 
role for gauge coupling unification~\cite{dienes,bachas,ross,zoup},  
for neutrino mass generation~\cite{neutrino}, to provide possible alternative  
solutions to the hierarchy problem~\cite{hierarchy1}-\cite{ab}, 
and as a transmitter of supersymmetry breaking between different  
boundaries~\cite{peskin}. On the other hand theories with a TeV higher 
dimensional Planck scale predict modifications of gravitational  
measurements in the sub-millimeter range~\cite{milli}, while some features of  
TeV-scale quantum gravity theories have recently been worked out~\cite{q-grav}. 
Finally experimental detection of TeV radii in future 
accelerators has been proposed as an unambiguous signature of 
large extra dimensions~\cite{abq,aceler}. 
 
In the above theoretical constructions, where compactification 
scales are in the TeV range, it is therefore tempting to assume 
that both the compactification scale of large dimensions and the 
scale of supersymmetry breaking have a common origin. This is 
the case if supersymmetry is broken by a continuous 
compactification along the compact dimension by means of the 
so-called Scherk-Schwarz (SS) mechanism~\cite{ss,ssst}. The SS 
mechanism has been recently used to break supersymmetry in 
sectors with which we only share gravitational~\cite{aq2,emilian} interactions 
(gravity mediated scenarios) and also in sectors that share 
gauge interactions with the observable sector~\cite{amq}-\cite{adpq}  
(gauge mediated scenarios).  
 
In this paper we will restrict ourselves to the 
latter scenarios, where both the compactification and the 
supersymmetry breaking scales are in the TeV range and can then 
leave a characteristic signature in the present or next 
generation of high-energy colliders: these scenarios will in 
this way be testable in the near future. In particular we will 
concentrate in simple five-dimensional (5D) models where SS 
compactification acts on the fifth dimension, as those presented 
in Refs.~\cite{pq,adpq}, where the main features of electroweak 
and supersymmetry breaking already appear. In this sense our 
approach will be a bottom-up one, but keeping in mind that it 
might possibly appear in compactifications of some 
more fundamental higher dimensional theory. However, as we will 
see the resulting low energy theory will show very little 
sensitivity to the physics at the high scale (cutoff). 
 
The plan of this paper is as follows. In section 2 we will 
present the simplest MSSM extension in 5D, compactified on 
$S^1\big/\mathbb{Z}_2$, along the lines of the model analyzed in 
Refs.~\cite{pq,adpq}. All non-chiral matter (the gauge and Higgs 
sectors) will be placed on the bulk of the fifth dimension while 
chiral matter (chiral fermion supermultiplets) live on the 4D 
boundaries. The corresponding 5D and 4D tree-level lagrangians 
and their interactions are explicitly written in subsections 2.1 
and 2.2, including contact 
interactions, between fields in the bulk and the boundary, which 
are necessary for the consistency of the theory. Supersymmetry 
is broken by the SS mechanism in the bulk, using the $U(1)_R$ 
$R$-symmetry of the $N=2$ supersymmetry algebra. The resulting 
massless spectrum contains just the SM particle content plus the 
sfermions living on the boundary that do not have excitations 
along the fifth dimension and thus do not receive any mass from 
the SS mechanism. One-loop radiative corrections  
will be studied in section 3 for the bulk, subsection 3.1, and 
the boundary, subsection 3.2. In the bulk the one-loop effective 
potential, in the background of the Higgs field zero-mode,  
is computed and found to have a closed analytic form in terms of  
polylogarithm functions. The Higgs squared-mass at the origin is 
computed by diagrammatic methods and shown to agree with that 
obtained from the effective potential: it is positive definite 
in the minimal MSSM extension, which asks for radiative corrections 
on the boundary to trigger electroweak symmetry breaking (a 
common procedure in theories where supersymmetry breaking is 
gauge mediated to the squark and slepton sector). Radiative 
corrections on the boundary are computed in subsection 3.2 by 
using diagrammatic methods. Supersymmetry breaking is mediated by 
gauge and Yukawa interactions from the bulk to the boundary. 
Explicit expressions are given for soft-breaking terms.  
In 
particular soft masses for sfermions and trilinear soft-breaking 
couplings are computed. Due to the presence of the fifth 
dimension the breaking is extremely soft and does not depend at 
all on the details of the ultraviolet physics. The soft masses 
can then be predicted as a function of a unique $R$-charge. 
Electroweak 
symmetry breaking is analyzed in section 4. It is triggered (as 
in gauge mediation) by two-loop corrections induced on the Higgs 
mass at the origin from sfermion soft masses on the boundary. 
These corrections are numerically relevant, due to the smallness 
of the bulk-generated positive Higgs squared-mass, and must be the 
leading two-loop corrections to the effective potential. 
Minimization of the whole effective potential leads to 
electroweak symmetry breaking at the correct scale with a light 
SM-like Higgs and very heavy supersymmetric particles. The rough 
features of the mass spectrum are as follows: the Higgs mass is bounded  
by $\simlt$ 110 GeV, squarks and sleptons have masses $\sim$ 1 TeV while  
heavy Higgses, gauginos and Higgsinos weigh $\sim$ 10 TeV. The 
model does not have any $\mu$-problem in the sense that the SS 
mechanism provides an effective $\mu$-parameter $\sim$ $1/R$. 
It predicts a light Higgs, which means that it can be probed in 
LEP, and a right-handed slepton as the lightest supersymmetric 
particle (LSP) which can generate a cosmological problem unless 
$R$-parity is broken or there are light right-handed  
neutrinos~\cite{adpq,k}. 
Other alternative models with heavier Higgses and a neutralino 
as LSP have been discussed in section 5. The most obvious 
possibility is having the chiral and gauge sector living in the bulk of 
the fifth dimension and the Higgs sector on the 
boundaries. These models suffer from the $\mu$-problem and 
present some experimental peculiarities. Finally the issue of 
unification in the presence of SS supersymmetry breaking has 
been studied in section 6. We have shown that gauge coupling 
running for the theory with SS supersymmetry breaking proceeds 
as in supersymmetric theories. In particular we have shown that 
the dependence of the gauge coupling running on the SS breaking 
parameters is extremely tiny, and corresponds to one part in 
$10^7$. Section 7 contains our conclusions and appendices A and 
B are devoted to present some technical details corresponding to 
the problem of gauge-fixing in the 5D theory and the calculation 
of the effective potential.    
   
\section{Tree-level Lagrangian}   
In this section we present a simple model based on an $N=1$ 5D theory    
compactified on $S^1\big/\mathbb{Z}_2$. The gauge sector and the non-chiral   
matter live in the bulk of the fifth dimension. In the minimal model the   
non-chiral matter consists on the Higgs sector. In non-minimal models   
we could add any vector-like extra matter, as e.g. $SU(2)_L$ triplets or   
complete $SU(5)$ representations plus their anti-particles. In what follows   
we will restrict ourselves to the minimal model.     
$N=1$ 4D chiral multiplets, with well defined renormalizable interactions   
with themselves and with the bulk fields, live on the 4D boundaries   
(fixed points of the orbifold $S^1\big/\mathbb{Z}_2$). The reason for    
having the chiral   
matter on the boundary is not chirality, that can be preserved by the   
$\mathbb{Z}_2$-projection, but the fact that chiral fermions in the bulk   
would not receive any mass from the electroweak breaking mechanism due to   
the underlying $N=2$ 4D symmetry of the Lagrangian.

\subsection{Fields in the 5D bulk}   
   
The gauge fields in the 5D bulk belong to the vector supermultiplet   
$\mathbb{V}=g\mathbb{V}^\alpha T^\alpha$, where $T^\alpha$ are the   
generators in the adjoint representation of the gauge group    
$SU(3)\times SU(2)_L\times U(1)_Y$ and $g$ the corresponding gauge    
couplings. The on-shell field content of $\mathbb{V}$ is    
$\mathbb{V}=(V_M,\lambda^i,\Sigma)$, where $M=\mu,5$ ($\mu$ is the 4D index),    
$\lambda^i$ is a simplectic-Majorana spinor whose superindex $i=1,2$ transforms   
as a doublet of the $SU(2)_R$ $R$-symmetry and $\Sigma$ is a real scalar.   
The Higgs fields belong to the hypermultiplets $\mathbb{H}^a$    
(an $SU(2)_L$-doublet with hypercharge $Y=-1/2$) whose superindex $a=1,2$    
transforms as the doublet of a global group $SU(2)_H$. The field content,    
on-shell, of the Higgs hypermultiplets is $\mathbb{H}^a=(H_i^a,\Psi^a)$,    
where $H_i^a$ are complex Higgs doublets and $\Psi^a$ Dirac spinors.   
   
The 5D lagrangian for vector and hypermultiplets is given by~\cite{pq,sohnius}  
\begin{eqnarray}   
{\cal L}_5&=&\Tr\frac{1}{g^2}\left\{-\frac{1}{2}F^2_{M,N}+   
\left|D_M\Sigma\right|^2+i\lambda_i\gamma^M D_M\lambda_i   
-\overline{\lambda}_i\left[\Sigma,\lambda^i\right]\right\}\nonumber\\   
&+&\left|D_MH^a_i\right|^2+i\overline{\Psi}_a\gamma^MD_M\Psi^a   
-\left(i\sqrt{2}H^{\dagger i}_a \overline{\lambda}_i\Psi^a+h.c.\right)   
-\overline{\Psi}_a\Sigma\Psi^a \nonumber\\   
&-&H^{\dagger i}_a \Sigma^2 H^a_i -\frac{g^2}{2}   
\sum_{m,\alpha}\left[H^{\dagger i}_a \left(\sigma^m\right)^j_i   
T^{\alpha} H^a_j \right]^2\, ,  
\label{lagVH}   
\end{eqnarray}   
where the $SU(2)_R\times SU(2)_H$ invariance is explicit.   
   
Upon compactification of the theory on $S^1\big/\mathbb{Z}_2$   
the 5D fields are classified under the $\mathbb{Z}_2$ parity   
into even and odd fields. The even fields are the vector   
multiplet $(V_\mu,\lambda_L^1)$ and the chiral multiplets   
$(H^2_2,\Psi^2_L)$ and $(H^1_1,\Psi_R^1)$. The odd fields are   
the chiral multiplets $(V_5,\Sigma,\lambda_L^2)$,   
$(H^2_1,\Psi_R^2)$, and $(H^1_2,\Psi_L^1)$. The $\mathbb{Z}_2$   
parity projects out half of the states. In fact the odd   
component of the zero modes is projected away while the net   
number of towers is divided by two since the positive non-zero modes of   
odd fields replace the negative non-zero modes of even fields.   
In this way the zero modes have an $N=1$ supersymmetry while the   
non-zero modes are arranged into $N=2$ multiplets in the way   
that can be seen in Table \ref{modes}.   
\begin{table}[H]   
\centering   
\begin{tabular}{|c|c|c|c|c|c|}   
\hline   
\multicolumn{3}{|c|} {$N=1$ zero modes} &    
\multicolumn{3}{c|} {$N=2$ non-zero ($n>0$) modes}\\   
\hline   
\emph {Vector} & \multicolumn{2}{c|} {\emph{Chiral}}& \emph{Vector} &   
\multicolumn{2}{c|} {\emph {Hyper}} \\   
\hline   
$V_\mu^{(0)}$ & $H_2^{2\,(0)}$ & $H_1^{1\,(0)}$ &   
$V_\mu^{(n)};\, \Sigma^{(n)}, V_5^{(n)}$  &   
$H_1^{1\,(n)};\, H_2^{1\, (n)}$ &   
$H_2^{2\,(n)};\, H_1^{2\, (n)}$\\   
$\lambda_L^{1\,(0)}$ & $\Psi_L^{2\, (0)}$ &   
$\Psi_R^{1\, (0)}$ & $\lambda_L^{1\,(n)};\,\lambda_L^{2\,(n)}$ &   
$\Psi_R^{1\, (n)};\,\Psi_L^{1\, (n)}$ &   
$\Psi_L^{2\, (n)};\, \Psi_R^{2\, (n)}$ \\   
\hline   
\end{tabular}   
\caption{Complete $N=1$ towers of states in the model.}   
\label{modes}   
\end{table}   
   
We have separated by a semicolumn the non-zero modes coming   
from an even 5D field (on the left hand side) from those coming from an   
odd 5D field (on the right hand side). We can then see in Table   
\ref{modes} how the complete ($n=-\infty,\dots,+\infty$) towers   
are constituted.   
   
As we can see from Table \ref{modes} the massless sector of the   
theory coincides with the MSSM. The zero modes of the gauge and Higgs 5D   
fields are 4D fields with $N=1$ supersymmetric interactions,    
while the chiral fields are required to live on the boundaries    
and so they are genuine 4D $N=1$ fields, as we will see in the next section,    
and they complete, along with the zero modes of the bulk fields, the MSSM.   
Of course, to agree with experimental data supersymmetry has to   
be broken. The Scherk-Schwarz (SS) mechanism~\cite{ss} was used   
in Ref.~\cite{pq} to break supersymmetry by means of a   
$U(1)_R\times U(1)_H$ global symmetry of the theory (a subgroup   
of the previously mentioned $SU(2)_R\times SU(2)_H$ group) with   
the corresponding charges $(q_R,q_H)$~\footnote{The SS   
mechanism has deep roots in supergravity~\cite{ss} and   
superstring~\cite{ssst}   
theories, where it is known to break spontaneously local    
supersymmetry. Had we included gravity, the 4D $N=1$   
supergravity constituted by the zero modes would be   
spontaneously broken with a gravitino mass $m_{3/2}=q_R/R$, where   
$R$ is the fifth dimension radius.}, which the mass spectrum   
depends upon~\footnote{Strictly speaking only the $R$-symmetry  
$U(1)_R$ breaks supersymmetry, while $U(1)_H$ generates a common  
(supersymmetric) mass shift for bosons and fermions. This can be  
explicitly seen from the mass spectrum of Table~\ref{masas}:   
had we taken $q_R=0$ the corresponding mass spectrum would  
be supersymmetric.}.    
   
After the SS supersymmetry breaking the $n$-KK mass eigenstates   
($n>0$) are now given by two Majorana fermions   
(gauginos) $\lambda^{(\pm n)}$,   
two Dirac fermions (Higgsinos)  $\widetilde{H}^{(\pm n)}$ and four   
scalar bosons (Higgses) $h^{(\pm n)}$ and $H^{(\pm n)}$, defined by   
\begin{eqnarray}   
\lambda^{(\pm n)}&\equiv&\left(\lambda_L^{1\,(n)}\pm\lambda_L^{2\,(n)}   
\right)\big/\sqrt{2}   
\nonumber\, ,\\   
\widetilde{H}^{(\pm n)}&\equiv&\left(\Psi^{1\,(n)}\pm\Psi^{2\,(n)}   
\right)\big/\sqrt{2}\nonumber\, ,\\   
h^{(\pm n)}&\equiv&\left[H^{1\,(n)}_1+H^{2\,(n)}_2\,\mp\left(H^{1\,(n)}_2  
-H^{2\,(n)}_1\right)\right]\big/2\nonumber\, ,\\  
H^{(\pm n)}&\equiv&\left[H^{1\,(n)}_1-H^{2\,(n)}_2\,\mp\left(H^{1\,(n)}_2  
+H^{2\,(n)}_1\right)\right]\big/2\, . 
\label{eigenst}   
\end{eqnarray}   
The corresponding masses are given in Table \ref{masas}   
\begin{table}[H]   
\centering   
\begin{tabular}{|c|c|}   
\hline   
Field & Mass \\   
\hline   
$\lambda^{(n)}$ & $\dfrac{|n+q_R|}{R}$ \\   
$\widetilde{H}^{(n)}$ & $\dfrac{|n+q_H|}{R}$ \\   
$h^{(n)}$ & $\dfrac{ |n+(q_R-q_H)|}{R}$ \\   
$H^{(n)}$ & $\dfrac{ |n+(q_R+q_H)|}{R}$ \\   
\hline   
\end{tabular}   
\caption{Mass eigenvalues of KK states.}   
\label{masas}   
\end{table}   
\noindent   
where now $n\in\mathbb{Z}$ runs over a whole tower.    
   
The masses of the zero modes of the even 5D Higgses   
\begin{eqnarray}  
h^{(0)} & = & \frac{1}{\sqrt{2}}\left(H^{1\, (0)}_1 
+H^{2\, (0)}_2\right)\, , \nonumber\\  
H^{(0)} & = & \frac{1}{\sqrt{2}}\left(H^{1\, (0)}_1-H^{2\, (0)} 
_2\right)\, ,   
\label{hH}  
\end{eqnarray}  
are given by $(q_R-q_H)/R$ and $(q_R+q_H)/R$, respectively.    
In the generic case $q_R\neq q_H$ no massless zero mode is left   
in the theory and this prevents electroweak symmetry breaking (EWSB).   
In the particular case $q_R=q_H\equiv \omega$ the Higgs doublet   
$h^{(0)}$ is massless, the Higgs sector coincides with that of the   
Standard Model and EWSB can proceed by radiative corrections, as   
we will discuss in section 4 where we will focus on this case.   
In the limiting case $\omega=1/2$ there is an extra massless   
mode, $H^{(-1)}$ and the Higgs sector is identical to that of   
the MSSM. This case will also be discussed in section 4 where we   
will prove that it is phenomenologically unappealing    
due to the fact that the down fermion sector remains massless   
since the Higgs that couple to it do not get a VEV.

\subsection{Fields on the 4D boundary}   
   
When considering a 5D theory in an orbifold $S^1\big/\mathbb{Z}_2$   
coupled to two 4D boundaries, one has to be careful when dealing    
with the off-shell formulation of 5D supermultiplets.   
The reason being, as discussed in Ref.~\cite{peskin}, that on   
top of the even 5D fields, also the $\partial_5$ of odd 5D   
fields  couple to the boundary, since  they are part of 
the auxiliary fields of the corresponding $N=1$   
supersymmetry algebra.   
   
In the case of a vector multiplet in the bulk, the off-shell   
multiplet is obtained by adding an $SU(2)_R$ triplet of   
real-valued auxiliary fields $X^A$~\cite{sohnius}:   
$\mathbb{V}=(V_M,\lambda^i,\Sigma,X^A)$. Classification under    
$\mathbb{Z}_2$ yields even $(V_\mu,\lambda_L^1,X^3)$ and odd   
$(V_5,\Sigma,X^{1,2})$ vector superfields. On the boundary, it was shown in   
Ref.~\cite{peskin}, that the off-shell multiplet   
$(V_\mu,\lambda_L^1,D)$, with   
\begin{equation}   
D=X^3-\partial_5\Sigma\, ,   
\label{dfield}   
\end{equation}   
closes the $N=1$ supersymmetry algebra. The reason for that can be traced    
from the transformation law of the particular combination   
(\ref{dfield}): it transforms    
as a total derivative under a supersymmetry transformation, which is    
precisely what it is expected for a D-field.   
   
The same argument is valid in the case of a hypermultiplet in the    
bulk~\footnote{In this case two different auxiliary fields are needed    
because of the $SU(2)$ automorphism group of the supersymmetry algebra.}.    
The even auxiliary field get mixed with the odd field of the 5D hypermultiplet.   
In particular, the off-shell Higgs hypermultiplet is given by   
$\mathbb{H}^a=(H^a_i,\Psi^a,F^a_i)$, which splits into even,   
$(H^2_2,\Psi_L^2,F^2_2)$, $(H^1_1,\Psi_R^1,F^1_1)$, and odd,   
$(H^2_1,\Psi_R^2,F^2_1)$, $(H^1_2,\Psi_L^1,F^1_2)$, chiral superfields.   
On the boundary, the off-shell chiral supermultiplets are:    
\begin{eqnarray}   
{\cal H}_2&=& \left(H^2_2,\Psi^2_L,F_2\right)\nonumber\, ,\\   
{\cal H}_1&=& \left(H^{1\dagger}_1,\overline{\Psi}_R^1,F^{\dagger}_1 
\right)\, ,    
\label{higgses}   
\end{eqnarray}   
where the corresponding $F$-fields are given by   
\begin{eqnarray}   
F_2 &=& F^2_2-\partial_5 H^2_1\nonumber\, ,\\   
F_1 &=& F^1_1-\partial_5 H^1_2 \, ,  
\label{ffield}   
\end{eqnarray}   
%
%
  
The auxiliary $F$- and $D$-fields appear now in    
the 4D lagrangian of the boundary fields as in a normal $N=1$ lagrangian.   
When those fields are integrated out by means of their equations of motions,    
new interactions terms do appear. We will now use those results in our    
particular model, the full development can be seen in Ref.~\cite{peskin}.   
   
Let us consider a family of left- and right-handed quark superfields     
$(\widetilde Q,q_L)$, $(\widetilde U,u_R)$ and $(\widetilde D,d_R)$     
living on the boundary at $x_5=0$~\footnote{The choice of the   
boundary $x_5=0$ is completely general since we can always make   
a change of variables in the orbifold $x'_5=x_5-\pi R$   
interchanging both boundaries. Our only hypothesis is that all   
chiral matter is located on the same boundary. Models with   
matter on both boundaries would give rise to a different   
phenomenology.}.   
Since only the 5D even fields are non vanishing at $x_5=0$,   
the gauge and Higgs supermultiplets form $N=1$ supermultiplets    
on the boundary.    
   
The gauge superfield,    
($V_\mu,\lambda_L^1,X^3-\partial_5\Sigma$), couplings to the   
left-handed quark superfields are given in Ref.~\cite{sharpe,peskin}.  
After eliminating the auxiliary field $X^3$, we  
get~\footnote{The last term in Eq.~(\ref{gauge5}) comes from the  
interaction $H^{\dagger\,i}_a(\sigma^3)^j_i\,X^3 H^a_j$   
in the off-shell formulation of the 5D  
lagrangian. Notice that this term was absent from Ref.~\cite{peskin}   
because only gauge fields were supposed to live in the bulk therein.  
It will appear in general whenever there is matter in the bulk  
and in the boundary with common gauge interactions. In our case  
no such term will appear for $SU(3)_c$, while terms mixing  
$\widetilde{Q}$, $\widetilde{U}$, $\widetilde{D}$,  
$\widetilde{L}$ and $\widetilde{E}$ in the boundary  
with the Higgs sector in the bulk, as in Eq.~(\ref{gauge5}),  
will appear for the $SU(2)_L$ and $U(1)_Y$ gauge groups.}  
\begin{eqnarray}   
{\cal L}_5&=&\Bigg[|D_\mu\widetilde Q|^2+i\bar q_L\sigma^\mu D_\mu q_L-   
\sqrt{2}i(\widetilde Q^\dagger \lambda^1_L q_L+h.c.)\nonumber\\   
&-&\widetilde Q^\dagger(\partial_5\Sigma)\widetilde Q   
-\frac{g^2}{2}\sum_\alpha(\widetilde   
Q^\dagger T^\alpha\widetilde Q)^2\delta(x_5)\nonumber\\  
&-& g^2 \sum_\alpha (\widetilde{Q}^\dagger T^\alpha \widetilde{Q})  
(H^{\dagger\,i}_a\left(\sigma^3\right)^j_i T^\alpha H^a_j ) 
\Bigg]\delta(x_5)\, .  
\label{gauge5}   
\end{eqnarray}   
A similar expression to Eq.~(\ref{gauge5}) holds for the couplings to the    
right-handed superfields. After reducing the lagrangian (\ref{gauge5})    
to 4D, one has in the physical basis   
\begin{eqnarray}   
{\cal L}_4&=&   
|D_\mu \widetilde Q |^2   
+i\bar q_L \sigma^\mu D_\mu q_L-   
\sum^{\infty}_{n=-\infty}   
\sqrt{2}   
i(\widetilde Q^\dagger \lambda^{(n)} q_L+h.c.)\nonumber\\   
&-&   
\sqrt{2}\sum^{\infty}_{n=1}   
\frac{n}{R}\widetilde Q^\dagger\Sigma^{(n)}\widetilde Q   
-\frac{g^2\pi R}{2}   
\sum_\alpha(\widetilde Q^\dagger T^\alpha\widetilde  
Q)^2\delta(0)\nonumber\\   
&-& \frac{g^2}{2}\sum_{n=-\infty}^{\infty}\,\sum_\alpha  
\widetilde{Q}^{\dagger} T^\alpha  
\widetilde{Q}\left(h^{(n)} T^\alpha H^{(n)\,\dagger} +\, h.c.\right)\, ,  
\label{gauge4}   
\end{eqnarray}   
where    
\begin{eqnarray}   
D_\mu&=&\partial_\mu+iV^{(0)}_\mu+   
i\sqrt{2}\sum^{\infty}_{n=1}V^{(n)}_\mu\, ,\nonumber\\   
\delta(0)&=&\frac{1}{\pi R}\sum^{\infty}_{n=-\infty} 1\, .   
\label{delta0}   
\end{eqnarray}   
and $g$ is now the 4D gauge coupling that is related to the 5D gauge 
coupling   
by $g|_{4D}=g|_{5D}\sqrt{\pi R}$.   
   
The Yukawa couplings of the Higgs chiral superfield ${\cal H}_2$ to   
the quark superfields on the boundary are given by    
\begin{eqnarray}   
{\cal L}_5&=&\Bigg[h_t\Big[H^2_2 q_L u_R+  
\Psi^2_L(\widetilde Q  u_R+ q_L\widetilde U) 
-(\partial_5 H^2_1)  \,\widetilde Q  \widetilde U 
+h.c.\Big]\nonumber\\   
&-&    
|h_t H^2_2\widetilde Q|^2-|h_t H^2_2\widetilde U|^2   
-|h_t\widetilde U \widetilde Q|^2\delta(x_5)\Bigg]\delta(x_5)\, ,    
\label{yuk5}   
\end{eqnarray}   
and a similar expression holds for the coupling of the Higgs   
multiplet ${\cal H}_1$ to the quark superfields on the boundary with   
the Yukawa coupling $h_b$.   
   
After reduction to 4D, the lagrangian in Eq.~(\ref{yuk5}) yields, in     
the physical basis,   
\begin{eqnarray}   
{\cal L}_4&=&   
\sum^{\infty}_{n=-\infty}   
\Bigg[\frac{h_t}{\sqrt{2}}\left(h^{(n)}-H^{(n)}\right) q_L u_R+   
h_t\widetilde H^{(n)}_L(\widetilde Q  u_R  
+ q_L\widetilde U)+h.c.\nonumber\\   
&-&    
\frac{h_t^2}{2}\Big[\left|\left(h^{(n)}-H^{(n)}\right)\widetilde  
Q\right|^2    
+\left|\left(h^{(n)}-H^{(n)}\right)\widetilde U\right|^2   
+\left|\widetilde Q \widetilde U\right|^2\Big]\nonumber\\   
&-&h_t\,\frac{n+q_R-q_H}{\sqrt{2} R}  
h^{(n)}\, \widetilde Q\widetilde U   
+h_t\,\frac{n+q_R+q_H}{\sqrt{2} R}  
 H^{(n)}\, \widetilde Q\widetilde U+h.c.\Bigg]\, .   
\label{yuk4}   
\end{eqnarray}   
For the bottom  sector one obtains  
\begin{eqnarray}   
{\cal L}_4&=&   
\sum^{\infty}_{n=-\infty}   
\Bigg[\frac{h_b}{\sqrt{2}}\left(h^{(n)}+H^{(n)}\right)^{\dagger} q_L d_R+   
h_b\overline{\widetilde H}^{(n)}_R(\widetilde Q  d_R  
+ q_L\widetilde D)+h.c.\nonumber\\   
&-&    
\frac{h_b^2}{2}\Big[\left|\left(h^{(n)}+H^{(n)}\right)\widetilde  
Q\right|^2    
+\left|\left(h^{(n)}+H^{(n)}\right)\widetilde D\right|^2   
+\left|\widetilde Q \widetilde D\right|^2\Big]\nonumber\\   
&+&h_b\,\frac{n+q_R-q_H}{\sqrt{2} R}  
h^{(n)\dagger}\, \widetilde Q\widetilde D  
+h_b\,\frac{n+q_R+q_H}{{\sqrt{2}} R}  
 H^{(n)\dagger}\, \widetilde Q\widetilde D+h.c.   
\Bigg]\, .   
\label{yuk4b}   
\end{eqnarray}   
   
\section{One-loop corrections}   
   
In the previous section we have depicted the tree-level   
structure of the model. Upon compactification to 4D on   
$S^1/\mathbb{Z}_2$ and supersymmetry breaking by the SS   
mechanism, the mass spectrum and couplings of zero-modes and   
KK-excitations depend on two parameters $q_R$ and $q_H$. Even if   
supersymmetry is broken in the bulk, for $q_R=q_H=\omega$ there   
is a massless Higgs doublet $h^{(0)}$, and EWSB should proceed by radiative   
corrections~\footnote{\label{nota} For the case $q_R\neq q_H$ all Higgs   
doublets acquire a tree level mass $\sim 1/R$, which would   
prevent the possibility of EWSB for values of $1/R$ in the TeV   
range.}. To achieve this task we have to compute the Higgs mass   
induced by one-loop radiative corrections in the bulk or, more   
generally, the one-loop effective potential in the bulk in the   
presence of a constant background Higgs field. On the other   
hand, supersymmetry, though broken in the bulk by the SS   
mechanism, is unbroken on the boundary. Transmission of   
supersymmetry breaking from the bulk to the boundary should   
proceed by radiative corrections as we will see in this section.   
This transmission will be gauge mediated and thus free of any   
problem related to flavour changing neutral currents. These   
issues will be studied in the present section.   
   
\subsection{Radiative corrections in the 5D bulk}   
   
We will start by considering the effect of a tower   
of KK-states with different masses for bosons and fermions   
\begin{eqnarray}   
m^2_B&=&{(n+q_B)^2}{\frac{1}{R^2}}\ ,\nonumber\\   
m^2_F&=&{(n+q_F)^2}{\frac{1}{R^2}}\  ,\ \ \ n=0,\pm 1,\pm 2,...   
\label{splitting}   
\end{eqnarray}   
We want to compute the one-loop effective potential for a massless    
scalar mode, $\phi$ (the one to be associated with the SM Higgs field   
$h^{(0)}$),  
induced by this tower of KK-states. This is given in the Landau    
gauge (see Appendix A) by   
\begin{equation}   
V=\frac{1}{2}   
{\rm Tr}\int\frac{d^4 p}{(2 \pi)^4}   
\sum^{\infty}_{n=-\infty}   
\ln\left (\frac{p^2+M^2(\phi)+\dfrac{(n+q_B)^2}{R^2}}   
{ p^2+M^2(\phi)+\dfrac{(n+q_F)^2}{R^2}}\right )\, ,   
\label{cw}   
\end{equation}   
where Tr is the trace over the number of  
degrees of freedom of the KK-tower    
and ${M^2(\phi)}$ is the $\phi$-dependent mass of the KK-states.   
In Eq.~(\ref{cw})  
we must first perform the summation over the KK-states and then integrate    
with respect to the momentum.  
The calculation  
has been performed in Appendix B using    
techniques borrowed from finite temperature calculations.    
We obtain   
\begin{equation} 
V=\frac{1}{128\pi^6 R^4}{\rm Tr}\Big[V(r_F,\phi)-V(r_B,\phi)\Big]\, ,
\label{pot}
\end{equation}   
where   
\begin{equation}   
V(r_i,\phi)=x^2 Li_3(r_i e^{-x})+3x Li_4(r_i e^{-x})
+3Li_5(r_i e^{-x})+h.c.\, ,
\label{pot2}
\end{equation}
\begin{equation}   
x=2\pi R\sqrt{M^2(\phi)}\, ,\ \ \ \    
r_i=e^{i2\pi q_i}
\, ,   
\label{couplings}   
\end{equation}   
and $Li_n(x)$ are the   
polylogarithm functions    
$$Li_n(x)=\sum^{\infty}_{k=1}\frac{x^k}{k^n}.$$   
As in finite temperature,    
the result is    
independent of the ultraviolet cutoff.    
The above potential is monotonically decreasing (increasing)  with $x$  
if $q_F<q_B\leq 1/2$ ($q_B<q_F\leq 1/2$);  
therefore if only Eq.~(\ref{pot}) is present, $\phi$   
is driven to zero (infinity). We can expand Eq.~(\ref{pot2})    
for $\phi\ll 1/R$ ($x\ll 1$)   
\begin{eqnarray}   
V(r,\phi)&=&   
3 [Li_5(r)+Li_5(r^*)]   
-\frac{x^2}{2}[Li_3(r)+Li_3(r^*)]\nonumber\nonumber\\   
&-&   
\frac{x^4}{8}\ln\frac{(1 - r)^2}{-r}  +    
\frac{1}{15} x^5+ {\cal O}(x^6)\, .   
\label{expanding}   
\end{eqnarray}   
This expansion is only valid if $r$ is not close to $1$,    
for $r=1$, the expansion is 
\begin{equation}   
V(r=1,\phi)=6\zeta (5)-\zeta (3) x^2+\left (\frac{3}{16}-   
\frac{1}{4}\log x\right )   
x^4+\frac{1}{15}x^5+{\cal O}(x^6)\, , 
\label{expanding1}   
\end{equation}   
where $\zeta(x)$ is the Riemann-zeta function.   
Notice that the only   
odd-term in the $x$-power expansion of the potential, $x^5$,   
cancels (see also Appendix B) in (\ref{pot}).    
This means that a cosmological phase   
transition in the 5D theory at temperatures $T>1/R$, that can be   
described by means of a genuine 5D field theory at finite temperature,   
is always second order. A similar observation has been recently   
done in Ref.~\cite{ddgr}.   
 
\begin{figure}[ht]   
\centering   
\epsfig{file=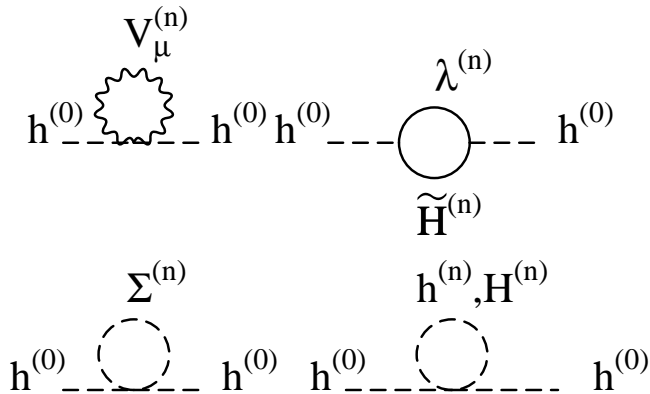,width=0.6\linewidth}   
\caption{Diagrams that contribute to the squared-mass of the   
Higgs field $h^{(0)}$.}   
\label{higgs}   
\end{figure}

From Eqs.~(\ref{pot}) and (\ref{expanding}), we can obtain  
the   mass of $\phi$ at the one-loop: 
\begin{equation}   
m^2_\phi=   
\frac{1}{32\pi^4}{\rm Tr}   
\left[\Delta m^2(q_B)-\Delta m^2(q_F)\right]   
\left.\frac{dM^2(\phi)}{d|\phi|^2}\right|_{\phi=0}     
\label{kkcont2}\, ,  
\end{equation}   
where   
\begin{equation}   
\Delta m^2(q)=\frac{1}{2R^2}\left[Li_3(e^{i2\pi q})+Li_3(e^{-i2\pi q})   
\right]\, \, .   
\label{lit}   
\end{equation}   
This coincides with the result in Ref.~\cite{adpq}, and can be interpreted    
diagrammatically in terms of the diagrams of Fig.~\ref{higgs}. 
In particular, we can calculate the one-loop mass of $h^{(0)}$ 
for the model described in the previous section 
in the case $q_R=q_H=\omega$. 
Considering only  the $SU(2)_L$ interactions, 
Eq.~(\ref{kkcont2})  yields  
\begin{equation}   
m^2_\phi=\frac{g^2_2}{64\pi^4}\left[9\Delta m^2(0)+3\Delta   
m^2(2\omega)- 12\Delta m^2(\omega)\right]\ .   
\label{m2fi}   
\end{equation}   
The Higgs squared-mass at the origin, defined by Eq.~(\ref{m2fi}),   
is positive definite and therefore radiative corrections on the boundary    
will be required to trigger EWSB, as we will   
see in the next sections. This procedure is a common one in   
theories where supersymmetry breaking is gauge-mediated to the   
sector of squarks and sleptons. The value of $m_\phi$ defined by   
 Eq.~(\ref{m2fi}) is a monotonically increasing function of $\omega$   
and takes values in the range $0<m_\phi<4\times 10^{-2}/R$ for   
$0<\omega<1/2$. Thus the scalar remains around two orders of   
magnitude lighter than the compactification scale.   
   
\subsection{Radiative corrections on the 4D boundary}   
   
The scalar fields on the boundary (i.e. squarks and sleptons)    
are massless at tree-level. Nevertheless, since supersymmetry is broken    
in the 5D bulk, the breaking will be transmitted    
to the fields on the boundary at the quantum level.   
   
\begin{figure}[ht]   
\centering   
\epsfig{file=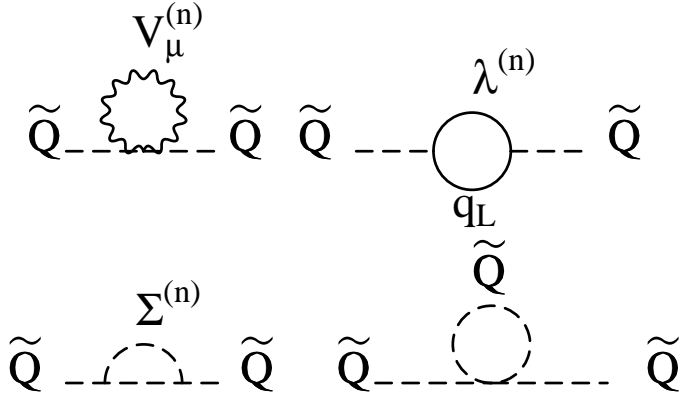,width=.6\linewidth}   
\caption{Diagrams that contribute to the mass of    
$\widetilde Q$ from the gauge sector.}   
\label{stop}   
\end{figure}

Let us consider the gauge corrections to the mass of $\widetilde Q$.   
The interactions between $\widetilde{Q}$ and the gauge   
supermultiplet are given in Eq.~(\ref{gauge4}). At the one-loop level,    
the diagrams that contribute to $m_{\widetilde Q}$    
are given in Fig.~\ref{stop}.   
The first (last) two diagrams are provided by the interactions in the  
first (second) line of Eq.~(\ref{gauge4}), while no contribution  
at one-loop comes from the terms in the third line of Eq.~(\ref{gauge4}).  
    
We must sum over the full tower of KK excitations.   
Using the methods of Appendix B to sum over KK-states, the   
diagrams of Fig.~\ref{stop} give   
\begin{equation}   
m^2_{\widetilde Q}=   
\frac{g^2C_2(Q)}{4\pi^4}   
\left[\Delta m^2(0)-\Delta m^2(q_R)\right]\, ,     
\label{gaugecont}   
\end{equation}   
where $C_2(Q)$ is the quadratic Casimir of the $Q$-representation   
under the gauge group~\footnote{We are using the convention for   
the generators: Tr$\, T^\alpha_R   
T^\beta_R=T(R)\delta^{\alpha\beta}$ and $T^\alpha_R   
T^\alpha_R=C_2(R)\cdot \boldsymbol{1}$, where $R$ is a  
representation of the gauge group and the unit matrix   
$\boldsymbol{1}$ has dimension $d(R)\times d(R)$, where $d(R)$ is   
the dimensionality of $R$. In particular if $\underline{N}$ is   
the fundamental representation of $SU(N)$,   
$T(\underline{N})=1/2$ and $C_2(\underline{N})=(N^2-1)/(2N)$,   
and for the adjoint (Adj) representation, $T(Adj)=C_2(Adj)=N$.},    
and $\Delta m^2(q)$ is given in Eq.~(\ref{lit}).   
   
\begin{figure}[htb]   
\centering   
\epsfig{file=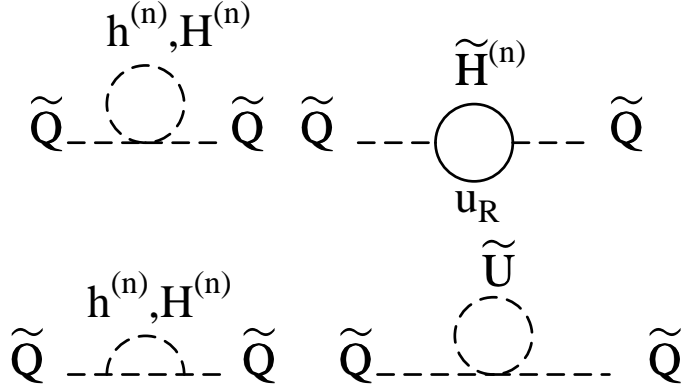,width=.6\linewidth}   
\caption{Diagrams that contribute to the mass of $\widetilde Q$    
from the Higgs sector.}   
\label{shiggs}   
\end{figure}

The interactions of $\widetilde{Q}$ with the Higgs sector can be   
read off from Eqs.~(\ref{yuk4}) and (\ref{yuk4b}). 
At the one-loop level the    
$h_t$-corrections to the mass of $\widetilde{Q}$ is provided by   
the diagrams of Fig.~\ref{shiggs}. The result is given by   
\begin{equation}   
m^2_{\widetilde Q}=   
\frac{h_t^2}{16\pi^4}   
\left[\Delta m^2(q_R+q_H)+\Delta m^2(q_R-q_H)   
-2\Delta m^2(q_{H})   
\right]\, .
\label{yukawacont}   
\end{equation}   
A similar expression to (\ref{yukawacont}) 
holds for the $h_b$-corrections.

In this way one can compute the radiative corrections to the   
masses of the sfermions $\widetilde{Q},\widetilde{U}$, $\widetilde{D}$,   
$\widetilde{L}$ and $\widetilde{E}$,   
in the model presented in section 2. The result   
can be written  as~\cite{adpq}:   
\begin{eqnarray}   
m_{\widetilde{Q}}^2 & = &   
\left(\frac{8}{6}\alpha_3+\frac{3}{4}\alpha_2   
+\frac{1}{60}\alpha_1\right) \Delta   
m_g^2+\frac{1}{2}(\alpha_t+\alpha_b)\Delta m_H^2\, , \nonumber\\   
m_{\widetilde{U}}^2 & = &   
\left(\frac{8}{6}\alpha_3+\frac{4}{15}\alpha_1\right) \Delta   
m_g^2+\alpha_t\Delta m_H^2 \, ,\nonumber\\   
m_{\widetilde{D}}^2 & = &   
\left(\frac{8}{6}\alpha_3+\frac{1}{15}\alpha_1\right) \Delta   
m_g^2+\alpha_b\Delta m_H^2 \, ,\nonumber\\   
m_{\widetilde{L}}^2 & = &   
\left(\frac{3}{4}  
\alpha_2   
 +\frac{3}{20}\alpha_1\right) \Delta m_g^2\, ,\nonumber\\   
m_{\widetilde E}^2 & = &   
\frac{3}{5}\alpha_1 \Delta m_g^2\, ,
\label{espectro}   
\end{eqnarray}   
where    
\begin{equation}   
\Delta m_g^2=\left[\Delta m^2(0)-\Delta m^2(q_R)\right]/\pi^3\, ,   
\label{mg2}   
\end{equation}   
and   
\begin{equation}   
\Delta m_H^2=\left[\Delta m^2(q_R+q_H)+\Delta m^2(q_R-q_H)-2   
\Delta m^2(q_H)\right]/2\pi^3\, ,   
\label{mh2}   
\end{equation}   
with $\Delta m^2(q)$ given in Eq.~(\ref{lit}). 
In Eq.~(\ref{espectro}) we have kept only the Yukawa couplings   
$h_{t,b}$ and defined $\alpha_{t,b}=h_{t,b}^2/4\pi$.   
   
\begin{figure}[ht]   
\centering   
\epsfig{file=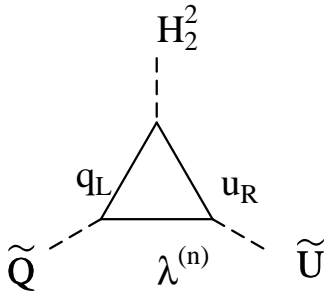,width=.3\linewidth}   
\caption{Diagrams that contribute to $A_t$.}   
\label{AtFig}   
\end{figure}

Finally we have computed the contribution of the KK-towers to   
the soft-breaking trilinear couplings between two boundary and   
one bulk fields. This contribution arises from gaugino loops as   
depicted in the diagram of Fig.~\ref{AtFig}. The leading   
contribution to the parameter $A_t$ is provided by the exchange   
of gluinos and given by, Ref.~\cite{adpq}:   
\begin{equation}   
A_t=\frac{8}{6}\frac{\alpha_3h_t}{2\pi^2}\left[i\,Li_2(e^{i2\pi q_R})-   
i\, Li_2(e^{-i2\pi q_R})\right]\frac{1}{R}\ .   
\label{at}   
\end{equation}   
The mixing $A_t$ vanishes at $q_R=0,1/2$ and is bounded by   
$|A_t|<2\times 10^{-2}/R$. The relative stop mixing is rather small:   
the function $a_t(q_R)\equiv A_t/m_{\widetilde{t}}$ is    
monotonically decreasing with $q_R$ and takes the values 0.33,   
0.18, 0.12 and zero for $q_R=1/10,\,1/4,\,1/3,$ and $1/2$,   
respectively.     
   
\section{Electroweak symmetry breaking}   
   
In section 2 we have described the tree-level mass spectrum of   
all KK-excitations. In particular we have seen that for   
$q_R=q_H\equiv \omega$ the zero mode  $h^{(0)}$ is massless   
and thus a good candidate to become the SM Higgs.   
It is useful to know how  $h^{(0)}$ is identified with  the two  
Higgses of the MSSM, $H_1$ and $H_2$. 
The latter are scalars with the SM quantum numbers  
$(\mathbf{1},\mathbf{2},1/2)$ and   $(\mathbf{1},\mathbf{2},-1/2)$, 
that couple respectively to the down and up fermion sector. 
Therefore, we have from eqs.~(\ref{yuk4}) and (\ref{yuk4b}) 
\begin{equation} 
h^{(0)}=\frac{1}{\sqrt{2}}\left(\sigma^2 H_1^* +H_2\right)\, , 
\label{mssm} 
\end{equation} 
where $\sigma^2$ acts on the $SU(2)_L$ indices 
of the Higgs doublet. 
In the range $0<\omega<1/2$ all modes $H^{(n)}$ get masses   
proportional to $1/R$ and become supermassive. In particular, the  
orthogonal field to (\ref{mssm}), that  
includes the MSSM Higgs $H^0$, $H^{\pm}$ and $A$, get a mass    
$2\omega/R$, and then a zero VEV. 
This implies 
$\langle H_2\rangle =   
\langle H_1\rangle$ or $\tan\beta=1$ in the MSSM language.  
On the other hand, the Higgsino zero mode also gets a   
tree-level mass $\omega/R$ as we have seen in Table \ref{masas}.   
Therefore, there is no $\mu$-problem  in this class of models.   
   
In order to find out whether there is a non-trivial    
minimum that can induce EWSB, the effective potential    
of Eq.~(\ref{pot}) has to be computed in the presence of the   
background field $\phi\equiv \sqrt{2}\langle h^{(0)} \rangle$.   
The effective potential in the bulk was given in Eq.~(\ref{pot2})    
and the background dependent masses which appear in   
Eq.~(\ref{couplings}) can be read off from the 5D lagrangian   
Eq.~(\ref{lagVH}). The mass of all KK-modes  as well as the   
corresponding number of degrees of freedom ($d_f$) coming from   
the SM group structure are displayed in Table \ref{masses-1},   
\begin{table}[ht]   
\centering   
\begin{tabular}{|c|c|l|}   
\hline   
Field & $d_f$ & $M^2(\phi)$ \\   
\hline   
$V^{(n)}_{\mu}$ & $3\times3$ &$\dfrac{n^2}{R^2}+\dfrac{1}{4}g^2_2\phi^2$ \\   
$\lambda^{(n)}$ & $3\times 2$ &$\dfrac{(n+\omega)^2}{R^2}+   
\dfrac{1}{4}g^2_2\phi^2$ \\   
$\widetilde{H}^{(n)}$ & $2 \times 2$ &    
$\dfrac{(n+\omega)^2}{R^2}+   
\,\dfrac{1}{2} C_2(H) g^2_2 \phi^2 $ \\   
$H^{(n)}$ &$3$ &   
$\dfrac{(n+2\omega)^2}{R^2}+\dfrac{1}{4} g^2_2 \phi^2 $ \\   
\hline   
\end{tabular}   
\caption{Degrees of freedom and background dependent masses for KK-modes.}   
\label{masses-1}   
\end{table}   
\noindent   
where $g_2$ is the $SU(2)_L$ gauge coupling.  
We are neglecting the  $U(1)_Y$ interactions.   
The counting of number of degrees of freedom is as follows.    
For the gauge fields we have three degrees of freedom coming   
from the $SU(2)_L$ structure (a triplet) and three from the trace   
over Lorentz indices in the Landau gauge. For the gauginos there   
are three degrees of freedom from the $SU(2)_L$ structure and two   
from the Majorana nature of gauginos. For the Higgsinos we have   
four degrees of freedom arising from their Dirac nature.   
Finally, in the Higgs sector, since there are no quartic   
couplings involving only the $h^{(n)}$ field, only some components of   
the $H^{(n)}$ field receive EWSB masses, in particular the charged   
components and the real part of the neutral component (the   
imaginary part does only receive the SS mass): a total of three   
degrees of freedom.   
   
By using the squared-mass and degrees of freedom values from Table    
\ref{masses-1} in  Eq.~(\ref{pot}) we see that the squared-mass term in   
the potential $m_\phi^2$ is positive for all values of   
$\omega$ (see Eq.~(\ref{m2fi}) and comments that follow it),   
which prevents the existence of a non-trivial EWSB minimum.   
However we have also to include in the one-loop potential the   
contribution from the 4D fields in the boundary. Technically   
speaking this is a two-loop contribution because the masses   
which appear there were generated at one-loop, Eq.~(\ref{espectro}).    
However this correction can be numerically   
relevant, due to the smallness of the bulk-generated mass at the   
origin for the Higgs field, and must be the leading two-loop   
correction to the effective potential. This contribution takes    
the usual form:   
\begin{equation}   
V_{4-D}=\frac{1}{64 \pi^2}{\rm Str}
M^4 \left ( \log \frac{M^2}{{\cal Q}^2}  
-\frac{3}{2}    
\right )\, ,   
\label{4-Dpot}   
\end{equation}   
where $M$ is the mass of the particle on the boundary which   
includes the EWSB contribution from the Higgs field $h^{(0)}$  
and ${\cal Q}$ is a    
renormalization scale. The only relevant contribution to   
Eq.~(\ref{4-Dpot}) is that coming from the top/stop    
sector~\footnote{We can neglect the sbottom sector, since we have   
$\tan\beta=1$, which would   
provide subdominant contributions.}. Now the one-loop potential   
(\ref{4-Dpot}) is similar to the MSSM one, once we have   
introduced the soft-breaking parameters from Eqs.~(\ref{espectro})    
and (\ref{at}). Finally we will choose the renormalization    
scale ${\cal Q}$   
as the boundary fixed by the lightest tree-level mass below which   
the theory can be considered 4D, i.e. the gaugino/Higgsino mass $\omega/R$.   
   
\begin{figure}[ht]   
\centering   
\epsfig{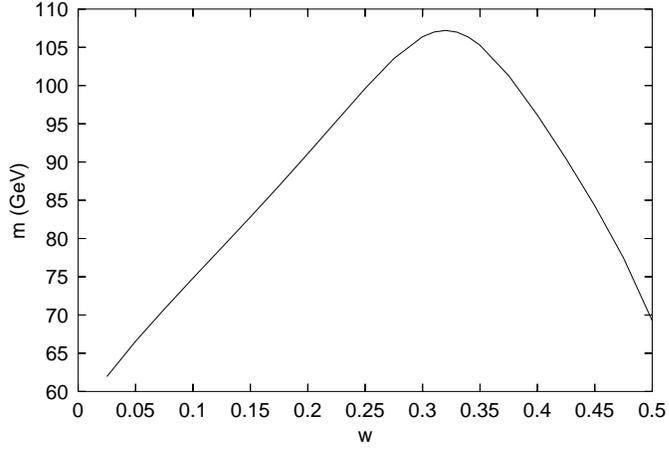}   
\caption{Mass of the Higgs field coming from the minimization   
of the effective potential.}   
\label{masashiggs}   
\end{figure}

\begin{figure}[H]   
\centering   
\epsfig{figure=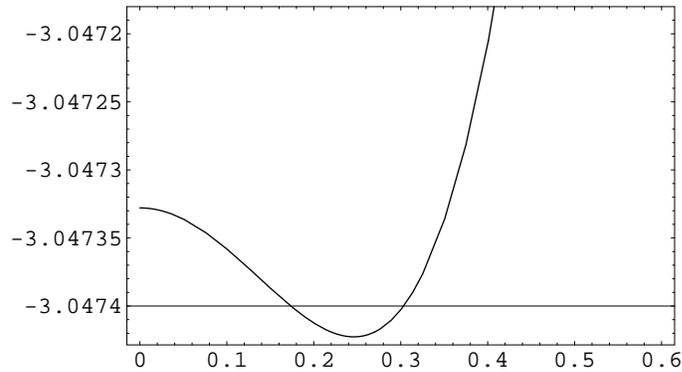,bbllx=88,bblly=15,bburx=396,bbury=192,clip=,width=.6\linewidth}   
\caption{Effective potential (in TeV$^4$) as a function of $\phi$   
(in TeV).}   
\label{pot_graf}   
\end{figure}

\begin{figure}[H]   
\centering   
\epsfig{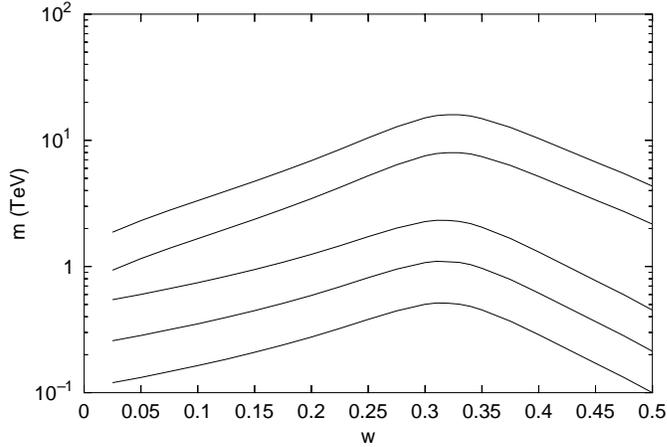}   
\caption{Masses of the different supersymmetric 
particles of the spectrum. From top to bottom we have the 
Higgs-doublet, the gauginos and Higgsinos, 
the squarks, the left-handed sleptons and the right-handed sleptons.}   
\label{spectrum}   
\end{figure}

Now we proceed in the following way. For a fixed value of $0<\omega<1/2$ we    
write the full effective potential, Eqs.~(\ref{pot}) plus (\ref{4-Dpot}), as    
a function of $\phi$ and $R$. By imposing the condition that $\langle\phi   
\rangle=246$ GeV we fix the value of the fifth dimension radius    
$R$~\footnote{We are not considering the compactification 
radius as a dynamical variable. 
Some ideas on how to dynamically fix the value of $R$ can be found in   
Ref.~\cite{adpq}.} as a function of $\omega$ and we can deduce from it the    
value of all soft-supersymmetry  
breaking parameters as well as the mass of the    
light physical Higgs $h^{(0)}$.  
The latter is shown in Fig.~\ref{masashiggs}    
as a function of $\omega$.   
We see from Fig.~\ref{masashiggs} that the maximum value is achieved when    
$\omega\sim 1/3$, and that for this particular value of $\omega$ the    
Higgs mass is $\sim\, 107$ GeV.   
The presence of a light Higgs is due to the fact that   
$h^{(0)}$  
is a flat direction of the $D$-term potential \cite{pq} and therefore  
its quartic coupling is zero at tree-level.   
The spectrum for $\omega=\frac{1}{3}$ 
is presented in Table \ref{quantities1}   
\begin{table}[H]   
\centering   
\begin{tabular}{|c|c|c|c|c|}   
\hline   
$1/R$&$m_{\lambda}$&$m_{\widetilde{Q}}$&$m_{\widetilde{L}}$&   
$m_{\widetilde{E}}$\\   
\hline   
 24 TeV & 8 TeV & 2.2 TeV & 1 TeV & 496 GeV\\   
\hline   
\end{tabular}   
\caption{Supersymmetric chiral spectrum for the model with    
$\omega=\frac{1}{3}$.}   
\label{quantities1}   
\end{table}   
\noindent   
for the supersymmetric chiral matter, and in Table \ref{quantities2}   
\begin{table}[H]   
\centering   
\begin{tabular}{|c|c|c|c|c|}   
\hline   
$\tan \beta$&$m_{\widetilde{H}}$&$m_{H^{\pm}}=m_{H^0}$&$m_A$&$m_h$\\   
\hline   
1 & 8 TeV & 15.9 TeV & 15.8 TeV & 107 GeV\\   
\hline   
\end{tabular}   
\caption{Higgs sector spectrum for the model with $\omega=\frac{1}{3}.$}   
\label{quantities2}   
\end{table}   
\noindent   
for the Higgs sector. For this case the shape of the effective potential   
is shown in Fig.~\ref{pot_graf}.   
We also plot in Fig.~\ref{spectrum} the mass spectrum of this    
model as a function of $\omega$. It is worth noting that the overall shape    
is similar in every case and also similar to the one in Fig.~\ref{masashiggs}.   
   
\bb{}

As can be seen in Tables \ref{quantities1}, \ref{quantities2} this model    
predicts a rather heavy spectrum of supersymmetric particles with the    
characteristic feature    
that the LSP is the right-handed selectron, due to the fact that supersymmetry    
breaking is transmitted via gauge interactions. Having the right-handed    
selectron as the LSP presents some cosmological difficulties. A possible way   
out is that R-parity is violated, so as the LSP is not stable. Other   
possibilities were discussed in Ref.~\cite{adpq}.   
We must remark that  the results on the light Higgs  
mass are quite sensitive to the value    
of ${\cal Q}$ in Eq.~(\ref{4-Dpot}); a little variation in this value,     
turns out to make a large variation in its mass.    
This implies that there is a certain uncertainty   
in the actual value of $m_h$, that would be aminorated   
if the two-loop contributions are incorporated.  
Nevertheless, we think that   
this two-loop calculation is not needed yet, since the   
experimental uncertainties on the values of $h_t$ and $\alpha_3$   
are still large. Furthermore we do not    
expect that the overall shape of the spectrum would be changed by these   
two-loop corrections in the bulk.    
  
Let us finally consider the case $\omega=1/2$. As we said,  
we have in this case two massless  
scalars in the  spectrum, $h^{(0)}$ and $H^{(-1)}$, that can be associated   
with the two Higgses of the MSSM:  
\begin{eqnarray}  
h^{(0)}&=&\frac{1}{\sqrt{2}}[\sigma^2 H_1^*+H_2]\nonumber\, ,\\  
H^{(-1)}&=&\frac{1}{\sqrt{2}}[\sigma^2 H_1^*-H_2]\, .  
\end{eqnarray}  
Nevertheless, this model is phenomenologically problematic  
since  in the limit $\omega=1/2$ the theory has a residual symmetry  
that does not allow for an $H_1H_2$ mixing term (i.e., $B\mu=0$ in the MSSM  
language) and therefore the VEV of $H_1$ is zero.  
To see that notice that for $\omega=1/2$  
the Higgsino $n$ KK-mode (in Table~\ref{masas}) is degenerated with the   
$-n-1$ KK-mode, and the $h^{(n)}$ KK-mode with the $H^{(-n-1)}$. 
Therefore 
the theory in the bulk and the boundary   
is invariant under the discrete transformation  
\begin{eqnarray}  
h^{(n)}&\longleftrightarrow &H^{(-n-1)}\, ,\nonumber\\  
\widetilde H_L^{(n)}&\longleftrightarrow &-\widetilde H_L^{(-n-1)} 
\nonumber\, ,\\ 
\widetilde H_R^{(n)}&\longleftrightarrow &\widetilde H_R^{(-n-1)} 
\nonumber\, ,\\  
(\widetilde Q,q_L)&\longleftrightarrow &(\widetilde Q,q_L) 
\nonumber\, ,\\  
(\widetilde U,u_R)&\longleftrightarrow &-(\widetilde U,u_R) 
\nonumber\, ,\\  
(\widetilde D,d_R)&\longleftrightarrow &(\widetilde D,d_R)\, .  
\label{pq1}  
\end{eqnarray} 
This symmetry implies $H_1\rightarrow H_1$ and $H_2\rightarrow -H_2$  
and consequently no mixing-mass term between the MSSM Higgses can  
be generated at any loop order.  
Note also that 
in the limit $\omega=1/2$  
the theory has also an $R$-symmetry,  
since the two Majorana gauginos   
$\lambda^{(n)}$ and $\lambda^{(-n-1)}$ can   
combine to form a Dirac fermion. 
Consequently the  trilinear terms (\ref{at}) are not generated in the  
limit $\omega=1/2$.  
  
\section{An alternative model}  
  
In this section we want to present a different possibility   
from the model studied above. We will assume   
quarks and leptons live in the   
bulk but the two Higgs supermultiplets live on the boundary.  
This possibility, although it suffers from the $\mu$-problem,  
presents a different and interesting phenomenology.

Quarks and leptons can arise from the zero modes of   
hypermultiplets. Compactifying in $S^
1/\mathbb{Z}_2$, we can obtain a   
chiral theory with $N=1$ supersymmetry \footnote{  
In a string theory, if the Higgs live on the boundary 
(twisted sector of the orbifold) and 
the quarks live in the bulk (untwisted sector), the 
Yukawa couplings can only be 
generated (to respect the $\mathbb{Z}_2$ symmetry of the orbifold) through
non-renormalizable couplings, as e.g., $XH_2q_Lu_R$, where $X$ is a SM-singlet
living on the boundary (twisted sector) which acquire a
VEV of the order of the high-scale $\Lambda$ (cutoff).}.
We can again use the SS-mechanism to break supersymmetry.   
Taking $q_R\not=0$, the squarks and sleptons get masses  
at the tree-level equal to $q_R/R$, and the  
massless sector in the bulk corresponds to  
the SM fermions and (as in the model above) to the SM gauge bosons.  
Now, however, since the two Higgs doublets live on the boundary,  
the scalar Higgs and Higgsinos are massless at tree-level.  
As in the model above, the scalar Higgses will get masses from   
their interactions with the bulk~\footnote{Notice that in this case,   
the finite one-loop contribution to the mass of $H_2$  
arising from the top/stop sector (that live in the bulk)  
can be negative (this is given by Eq.~(\ref{yukawacont})  
with $q_H=0$) and dominate over the positive gauge contribution  
(Eq.~(\ref{gaugecont})). This would make the EWSB easier and  
lead to a heavier lightest Higgs mass.} but the Higgsinos  
will remain massless at any loop order; the model  
suffers from the $\mu$-problem.  
To make the model phenomenological viable  
a Higgsino mass must be generated.  
A simple way to give mass to the Higgsino is through a nonzero VEV of   
a singlet $S$-field coupled to $H_1H_2$.  
We will not specify here how $S$ gets a VEV.  
We just want to point out that the phenomenology of this scenario  
is quite different from the previous one.  
The Higgsinos are the LSPs. Since the gaugino mass is very large,  
the charged Higgsino will be slightly  
heavier than the neutral Higgsino, with a mass difference of   
few GeVs \cite{gp}.   
Therefore the LSP is a neutral particle that does not present the  
cosmological problems of a charged one.  
The mass degeneracy of the Higgsinos, however, makes their detection   
problematic. The usual decay channel  
$\widetilde H^+\rightarrow \widetilde H^0 e\nu$ cannot be used to  
detect $\widetilde H^+$,  
because of the lack of energy of the  $e\nu$ decay products.  
In this case, the detection of the charged Higgsino must be carried out  
by  photon tagging.   
  
\section{Unification with SS supersymmetry breaking}  
  
The phenomenology of gauge coupling unification in the presence of extra  
dimensions was studied in Refs.~\cite{dienes} where it was proven that a  
sufficient condition for unification is that the ratio  
\begin{equation}  
R_{ij}=\frac{b_i^{\rm KK}-b_j^{\rm KK}}{b_i^{\rm MSSM}-b_j^{\rm MSSM}}\, ,  
\label{cond}  
\end{equation}  
does not depend on $(i,j)$,  
where $b_i^{\rm MSSM}$ are the MSSM $\beta$-function coefficients and   
$b_i^{\rm KK}$ those of the $N$=2 KK-excitations. Using  
$b^{\rm MSSM}=(33/5,1,-3)$ and $b^{\rm KK}=(6/5,-2,-6)$ for the model  
presented in section 2 and in Ref.~\cite{pq},   
we can see that it does not satisfy the  
necessary requirements to fulfill gauge coupling unification. This fact has  
recently motivated the suggestion of enlarging the model with the extra  
hypermultiplets $\mathbb{F}^a$ ($a=1,2$)~\cite{zurab} 
which are $SU(3)_c \times SU(2)_L$ singlets and  
having hypercharge $Y=1$. In Ref.~\cite{zurab} it was proven that the 
enlarged model unifies as well as the MSSM provided that we can introduce a   
supersymmetric mass term $\mu_F\simlt 1/R$ for the new fields, that can be
done by means of singlet fields getting non-zero VEVs.
In this section we will discuss  
how to incorporate the fields $\mathbb{F}^a$ in our formalism where the SS  
mechanism breaks supersymmetry and how the extra $\mu$-problem can be solved  
in a similar fashion as the $\mu$-problem for Higgs fields.  
Also we will study the issue of gauge  
coupling unification in the presence of SS supersymmetry breaking.  
  
Introduction of the hypermultiplets $\mathbb{F}^a$ in our formalism can  
be done  along the same lines as those leading to the mass spectrum and
 interactions of the hypermultiplets $\mathbb{H}^a$ 
in section 2.1. In fact, the 5D  
lagrangian for the vector multiplet ($\mathbb{V}$) and the hypermultiplets  
$\mathbb{F}^a$ is given by Eq.~(\ref{lagVH}) after replacing  
$\mathbb{H}^a\rightarrow\mathbb{F}^a$, yielding mass eigenvalues as those  
given in Table~\ref{masas} after the replacement $q_H\rightarrow q_F$. There  
is also the coupling of matter supermultiplets on the boundary with the  
hypermultiplets $\mathbb{F}^a$ corresponding to the last term of   
Eqs.~(\ref{gauge5}) and (\ref{gauge4}). Note that by choosing $q_F\neq q_R$  
no massless modes do appear in the spectrum and all zero modes   
will acquire masses $\simlt 1/R$, as required by gauge coupling unification,   
without any need to introduce supersymmetric mass terms.  
  
For scales ${\cal Q}\ll 1/R$ the standard Coleman-Weinberg prescription for  
the gauge couplings gives the one-loop result for the previous model,  
\begin{eqnarray}  
\alpha_i^{-1}({\cal Q})&=& \alpha_i^{-1}(\Lambda)+\frac{b_i^{\rm SM}}{2\pi}  
\ln\frac{M_c}{\cal Q}+\frac{b_i^{\rm MSSM}-b_i^{KK}}{2\pi}\ln\frac{\Lambda}  
{M_c}\nonumber\\  
&+&\frac{1}{2}\,\frac{1}{4\pi}\ \sum_{f\in  
\mathbb{V},\mathbb{H}^a,\mathbb{F}^a}b^{(f)}T(R_f)  
\ \sum_{n=-\infty}^{\infty}\,\int_{r\Lambda^{-2}}^{  
rM_c^{-2}}\frac{d\, t}{t}e^{-(n+\omega_f)^2t\, M_c^2}\, ,   
\label{colwein}  
\end{eqnarray}  
where the mass of the $n-th$ KK-excitation of the $f$-field is $(n+\omega_f)/R$,
we have already introduced the $\mathbb{Z}_2$ projection,  
$M_c=1/R$, the cutoff coefficient is~\cite{dienes} $r=\pi/4$,  
$b^{\rm SM}=(41/10,-19/6,-7)$ and the  
$\beta$-function coefficients $b^{(f)}$ are: $b$(gauge boson)=$-11/3$,   
$b$(Weyl fermion)=$2/3$  
and $b$(complex scalar)=$1/3$. They obviously satisfy the condition  
$\sum_f b^{(f)} T(R_f)=b_i^{\rm KK}$ where, for the enlarged MSSM model,  
\begin{equation}  
b^{\rm KK}=(18/5,-2,-6)\, .  
\label{coefic}  
\end{equation}  
  
The last integral in (\ref{colwein})  
\begin{equation}  
I(\omega)=\sum_{n=-\infty}^{\infty}  
\int_{r(M_c/\Lambda)^2}^{r}\frac{dx}{x}e^{-(n+\omega)^2 x}\, ,  
\label{integral}  
\end{equation}  
can be computed with the help of  
the Poisson resummation formula  
\begin{equation}  
\sum_{n=-\infty}^{\infty} e^{-(n+\omega)^2 x}=\sqrt{\frac{\pi}{x}}  
\sum_{n=-\infty}^{\infty} e^{-\frac{\pi^2}{x}n^2-2i\pi n\omega}\, ,  
\label{poisson}  
\end{equation}  
and it can be approximated by  
\begin{eqnarray}  
I(\omega)&=&4\Big(\frac{\Lambda}{M_c}-1+\frac{1}{2}\sum_{n=1}^{\infty}  
\frac{\cos(2\pi n\omega)}{n}\left[1-Erf(2n\sqrt{\pi})\right]\Big)\nonumber\\  
&\simeq&4\Big(\frac{\Lambda}{M_c}-1+\frac{1}{4\pi}e^{-4\pi}  
\cos(2\pi\omega)\Big)\, ,  
\label{I}  
\end{eqnarray}  
where $Erf$ is the error function and we have used its asymptotic expansion,  
which is dominated by  the $n=1$ mode. The function (\ref{I})  
exhibits a tiny $\omega$-dependence since $e^{-4\pi}/4\pi\sim 10^{-7}$.  
  
Therefore Eq.~(\ref{colwein}) looks like  
\begin{equation}  
\alpha_i^{-1}({\cal Q})= \alpha_i^{-1}(\Lambda)+\frac{b_i^{\rm SM}}{2\pi}  
\ln\frac{M_c}{\cal Q}+\frac{b_i^{\rm MSSM}-b_i^{KK}}{2\pi}\ln\frac{\Lambda}  
{M_c}+\frac{b_i^{\rm KK}}{2\pi}\left(\dfrac{\Lambda}{M_c}-1\right)\, ,  
\label{un}  
\end{equation}  
and unification proceeds, concerning KK-modes, as in the supersymmetric case.  
 
\section{Conclusions}   
 
In this paper we have addressed the issue of  
extra dimensions at the TeV-scale  
as a possible origin of electroweak breaking for the Standard Model, as well  
as the source of soft breaking terms in its supersymmetric extensions.  
In a bottom-up approach we have constructed a minimal extension of the MSSM  
in a five-dimensional space-time, with one ``large'' space dimension with 
radius $\sim$ 1 TeV$^{-1}$, which exhibits all the required features.    
 
After compactification on the segment $S^1\big/\mathbb{Z}_2$  
the zero modes of 
the model constitute the spectrum of the MSSM, while the  
non-zero Kaluza-Klein  
excitations are arranged into $N=2$ supermultiplets. When supersymmetry  
is broken by the 
Scherk-Schwarz mechanism, making use of the underlying 
$N=2$ $SU(2)_R$ algebra, the massless 
spectrum in the Higgs and gauge sector coincides with that of the pure 
Standard Model while their fermionic partners acquire tree-level masses. 
Chiral sfermions are massless at the tree-level, since chiral matter is 
supposed to live on the four dimensional boundary of the fifth dimension. 
 
Supersymmetry is gauge and Yukawa mediated 
to the chiral sector by radiative  
corrections, which also trigger electroweak symmetry breaking. 
In those aspects 
the model shares common features with 
any gauge-mediated supersymmetry breaking 
model but with a very characteristic spectrum. Electroweak 
breaking is achieved with a rather light SM Higgs (lighter than 
$\sim$ 110 GeV), very heavy gauginos, Higgsinos and non-SM 
Higgses (with masses $\sim 1/R$), and chiral sfermions at some 
intermediate squared-masses ($\sim \alpha_i/R^2$ where $\alpha_i$ are 
either gauge or Yukawa couplings. The model does not suffer from 
any $\mu$-problem since there is an effective $\mu$ parameter 
$\sim 1/R$: in fact both the Higgsinos and the pseudoscalar Higgs 
acquire masses $\sim 1/R$.  
 
Gauge coupling running and unification in the presence of extra dimensions  
is studied when supersymmetry is broken by the Scherk-Schwarz 
mechanism. Concerning the running, in the presence of the fifth 
dimension, we have computed the contribution from a tower of 
KK-excitations with a mass given by $(n+\omega)/R$. The leading 
contribution, $\sim(\Lambda R-1)$, is $\omega$-independent, 
while all the $\omega$-dependence is concentrated in sub-leading 
contributions which are corrections $\sim 
10^{-7}\cos(2\pi\omega)$, and therefore negligible.  
 
The model we have presented in this paper must not be considered 
as a unique model, with unique predictions, but rather as a 
representative of a class of models sharing common features: 
there is some extra dimension(s) at the TeV-scale which triggers, 
through the Scherk-Schwarz mechanism, the breaking of 
supersymmetry and the electroweak symmetry. Other possibilities, 
apart from the minimal model we have studied in great detail,  
have been pointed out, that 
offer a wide and rich variety of different phenomenological 
outcomes. Since the common feature of all these models is the 
appearance of extra dimensions at the TeV-scale, they are 
testable at present and future accelerators, which makes it worth 
pursuing their analysis and, in particular, the search of their experimental 
signatures. From a more fundamental point of view it would be 
important to find consistent string vacua reducing to our models 
at low energies: e.g. D4-branes in compactifications of Type IIA 
orientifolds or D5-branes in Type I (I') or Type IIB 
orientifolds. Some of these ideas are at present being  
investigated~\cite{TeV,zurab,TypeI}.

\section*{Acknowledgements} 
We thank Ignatios Antoniadis, Savas Dimopoulos and 
Zurab Kakushadze for discussions.
The work of AD was supported by the Spanish Education Office  
(MEC) under an \emph{FPI} scholarship.   
 
\newpage   
\appendix   
\section{Gauge Fixing}   
\bb{}   
In this appendix we will show that, in the 5D Landau gauge,   
the goldstone boson $V_5^{(n)}$ decouples from the lagrangian.    
This is a nice feature of the five dimensional theory.   
\bb{}   
   
We start with the kinetic     
gauge boson lagrangian plus a gauge fixing term~\footnote{We   
consider first an abelian theory, 
the translation to the non-abelian case being    
straightforward.} in five dimensions:   
\begin{equation}   
\mathcal{L}_5=-\frac{1}{4}F^{MN}F_{MN}-
\frac{1}{2\xi}(\partial_{M}V^{M})^2\, . \label{gaugelagran}   
\end{equation}      
Dimensional reduction of (\ref{gaugelagran}) leads to the    
following expression:   
\begin{eqnarray}   
\mathcal{L}_4&=&\sum_{n=0}^{\infty}\left (-\frac{1}{4}F^{(n)\mu\nu}   
F^{(n)}_{\mu\nu}+\frac{n^2}{2R^2}V^{(n)}_{\mu}V^{(n)\mu}-\frac{1}{2\xi}   
(\partial_{\mu}V^{(n)\mu})^2\right )\nonumber\\   
&+&\sum_{n=1}^{\infty}\left (\frac{1}{2}\partial_{\mu}V_5^{(n)}   
\partial^{\mu}V_5^{(n)}-\frac{n^2}{2\xi R^2}V_5^{(n)2}+   
\frac{n}{R}
\left(\frac{1}{\xi}-1\right)\partial_{\mu}V^{(n)\mu}V_5^{(n)}\right)\, .   
\label{gaugereduc}   
\end{eqnarray}   
The propagators for $V^{(n)}_{\mu}$ and $V^{(n)}_5$,    
using Eq.~(\ref{gaugereduc}), are given by
\begin{eqnarray}   
\langle V^{(n)}_{\mu}V^{(n)}_{\nu}\rangle &=&\frac{-i}{p^2-   
\frac{n^2}{R^2}}\left (\eta_{\mu\nu}+(\xi-1)\frac{p_{\mu}p_{\nu}}   
{p^2-\xi \frac{n^2}{R^2}}\right )\, ,\nonumber\\   
\langle V^{(n)}_5 V^{(n)}_5 \rangle&=&\frac{i}{p^2-\frac{n^2}{\xi R^2}}\, .   
\label{propagators}   
\end{eqnarray}   
Taking now the limit  $\xi\rightarrow 0$  in Eq.~(\ref{propagators})    
we obtain the normal propagator for a massive gauge boson in the Landau    
gauge, but we note that the propagator for the goldstone boson vanishes,    
so $V^{(n)}_5$ decouples from the lagrangian for each mode.    
The conclusion is that the Landau gauge in five dimensions leads    
to the Landau gauge in four dimensions for the gauge bosons and to   
the \emph{unitary gauge} for the goldstone bosons.

\section{The effective potential}   
   
In this appendix we will compute the effective potential   
corresponding to a tower of bosonic and fermionic KK-modes with   
masses given by Eq.~(\ref{splitting}). The basic integral we   
have to compute is then:   
\begin{equation}   
V=\frac{1}{2}\int \frac{d^4 p}{(2\pi)^4}\sum_{n=-\infty}^{\infty}   
\ln\left[\ell^2 E^2+(n+\omega)^2\pi^2\right]\, ,   
\label{inicio}   
\end{equation}   
where $\ell=\pi R$ is the length of the segment,    
$\omega$ is either $q_B$ or $q_F$ in (\ref{cw}), a global   
minus sign has to be added in the case of fermions, and    
$E^2=p^2+M^2(\phi)$.   
   
We will first evaluate the infinite sum over the KK-modes in    
\begin{equation}   
W=\frac{1}{2}\sum_{n=-\infty}^{\infty}\ln\left[(\ell\,E)^2+(n+\omega)^2   
\pi^2\right]\, ,   
\label{funW}   
\end{equation}   
or equivalently, in   
\begin{equation}   
\frac{\partial W}{\partial E}=\ell^2 E \sum_{n=-\infty}^{\infty}    
\frac{1}{(\ell\, E)^2+(n+\omega)^2\pi^2}\, ,   
\label{derW}   
\end{equation}   
by means of well known techniques used in field theory at finite   
temperature. To this end we will make use of the identity
\begin{equation}   
\frac{1}{(\ell\,E)^2+(n+\omega)^2\pi^2}=2i \,    
\lim_{z\rightarrow   
(n+\omega)\pi}\frac{z-(n+\omega)\pi}{e^{2iz}-e^{2i\omega\pi}}   
\frac{e^{2i\omega\pi}}{(\ell\,E)^2+z^2}\, ,    
\label{carajillo}   
\end{equation}   
and replace the infinite sum in (\ref{derW}) by an integral in   
the complex plane $z$ over a contour which is the sum of the   
contours encircling anti-clockwise the infinite number of poles   
along the real axis at $z=(n+\omega)\pi$. We now deform this   
contour into a contour going from $+\infty+i\epsilon$ to   
$-\infty+i\epsilon$ (which can be closed clockwise at    
$|z|\rightarrow\infty$) and a contour going from   
$-\infty-i\epsilon$ to $+\infty-i\epsilon$ (which can equally    
be closed clockwise at $|z|\rightarrow\infty$), and make use of   
the residues theorem to perform the integral over $z$. This   
easily yields,   
\begin{equation}   
W=\ell\, E+\frac{1}{2}\left\{\ln\left(1-r e^{-2\ell E}\right)+    
\ln\left(1-\frac{1}{r}e^{-2\ell E}\right)\right\}\ ,   
\label{Wsol}   
\end{equation}   
where $r=\exp(-2i\omega\pi)$,    
which corresponds to the decomposition of the effective potential as   
\begin{equation}   
V=V^{(\infty)}+V^{(R)}\, .   
\label{descomp}   
\end{equation}   
The first term in (\ref{descomp}) comes from the first term in   
(\ref{Wsol}) and provides a genuine 5D effective   
potential. It corresponds to the decompactification limit   
($R\rightarrow \infty$) of the theory and it is similar to the   
zero-temperature term which is obtained in field theory at   
finite temperature calculations. Since the 5D theory is not   
renormalizable, it must be computed by introducing a physical   
cutoff $\Lambda$ in the integral. Then the integral can be given   
an analytical form as   
\begin{eqnarray}   
V^{(\infty)}&=& \ell\int\frac{d^4 p}{(2\pi^4)}\sqrt{p^2+M^2(\phi)}   
\nonumber\\   
&=&\frac{\ell}{16\pi^2}\left\{ \frac{4}{15} M^5+\frac{2}{15}   
\sqrt{\Lambda^2+M^2}\left(3\Lambda^4+\Lambda^2   
M^2-2M^4\right)\right\} .   
\label{Vinfty}   
\end{eqnarray}   
Notice that upon expansion of (\ref{Vinfty}) in powers of $M$   
the only odd power is given by the $M^5$-term. This term cancels   
the similar one in the expansion of $V^{(R)}$ (see   
Eq.~(\ref{expanding})) and there is no odd-power term in the   
expansion of the effective potential~\footnote{This is in   
contradistinction with the case of a 4D theory at finite   
temperature where there is an $M^3$-term in the bosonic   
expansion, which triggers first-order phase transitions. As a   
consequence, in a 5D theory at finite temperature the phase   
transition should be of second-order.}. However, in a supersymmetric theory   
$V^{(\infty)}_B=V^{(\infty)}_F$ and the contributions to the   
effective potential from $V^{(\infty)}$ do cancel.   
The second term in (\ref{descomp}) comes from the last two terms   
in (\ref{Wsol}) which yields, upon integration over angular   
variables,    
\begin{equation}   
V^{(R)}=\frac{1}{32\pi^6 R^4}\int_0^\infty dy\ y\left[   
\ln\left(1-r e^{-2\sqrt{y+(\pi R M)^2}}\right)+\left(r\rightarrow   
\frac{1}{r}\right) \right]\ .   
\label{VR}   
\end{equation}   
Finally, the $y$-integral can be performed analytically giving   
the result that can be found in Eqs.~(\ref{pot}) and   
(\ref{pot2}).

\end{document}